\documentclass[letterpaper]{article}

\usepackage[T1]{fontenc}
\usepackage{amsmath,amsfonts,amssymb}

\usepackage{geometry}
\usepackage{setspace}

\usepackage[%
backend=biber,
style=phys,
giveninits=true,
biblabel=brackets,
]{biblatex}
\usepackage{graphicx}
\usepackage{float}
\newfloat{scheme}{htbp}{los}
\floatname{scheme}{Scheme}
\floatname{chart}{Chart}
\newfloat{graph}{htbp}{loh}

\usepackage{authblk}

\usepackage{xurl}
\usepackage{caption}
\usepackage{siunitx}
\usepackage{xcolor}
\usepackage[autostyle]{csquotes}
\usepackage{hyperref}
\usepackage[english]{babel}
\usepackage[english,noabbrev]{cleveref}

\DeclareSIUnit\angstrom{\text{\AA}}
\DeclareSIUnit\rydberg{Ry}
\DeclareSIUnit\bar{bar}
\DeclareUnicodeCharacter{0301}{$^\prime$}
\DeclareUnicodeCharacter{2032}{$^\prime$}

\title{Impact of the Lattice Constant on the Polymorphism of Organic/Inorganic Interfaces}
\author[1]{Christoph Wachter}
\author[1,*]{Oliver T. Hofmann}
\affil[1]{Institute of Solid State Physics, NAWI Graz, Graz University of Technology, Petersgasse 16, 8010 Graz, Austria}
\date{*Email: o.hofmann@tugraz.at}

\begin{document}
\maketitle

\begin{abstract}
The polymorphism of organic/metal interfaces influences many of their properties. As a result, a host of
contemporary research focuses on analyzing the factors which are pertinent for modifying polymorphism.
In this work, we elucidate how the lattice constant of the underlying lattice affects the energetic landscape of adsorbate monolayers for the model system of tetracyanoquinodimethane (TCNQ) on coinage metal surfaces with varying lattice constants.
In particular,
we focus on how the adsorbate-adsorbate and the adsorbate-substrate interaction are affected when increasing
the lattice constant and changing the surface chemistry.
Based on these investigations, we show that the adsorbate-substrate interaction for some adsorption geometries can change significantly with the lattice constant. In addition, due to a transition from repulsive to attractive adsorbate-adsorbate interactions, polymorphs with tight packing become more favorable, if the lattice constant is increased, resulting in a lattice-constant-based phase transition.
\end{abstract}

\section{Introduction}

The polymorphism of organic/metal interfaces reflects the intricate balance between the interactions
between adsorbate and substrate and between the adsorbate molecules themselves. Moreover, the polymorphism has
direct influence on many aspects of these interfaces, such as their electronic,\cite{ambrosch-draxlRolePolymorphismOrganic2009}
mechanical,\cite{upadhyayRelationshipCrystalStructure2013,wangRelationshipsCrystalStructures2017} optical,\cite{tiagoInitioCalculationElectronic2003} and
thermal\cite{davidStructureThermalProperties2012} properties.
For that reason, it is not only desirable to explore the polymorphism of a given interface, but also to attempt to manipulate it.
A key ingredient for influencing the polymorphism of the organic adlayer is the underlying substrate.\cite{papageorgiouChemicalTransformationsDrive2012}
Changing the substrate entails a change in the chemical environment and in reactivity, that affects preferred adsorption sites, adsorbate orientation and charge transfer into the adlayer.\cite{rochefortRoleStructureReactivity2021,romanerTheoreticalStudyPTCDA2009}
For the common fcc-metal surfaces a change in chemistry is usually accompanied by a change in the lattice constant of the substrate.
While more subtle, this change in lattice constant has also the potential to influence the adsorption geometries and polymorphism of the organic molecules.
However, it is difficult to tell whether chemistry, lattice constant or both are the deciding factor when modifying polymorphism by changing the underlying substrate.

In this work, we disentangle the effects of substrate lattice constant and chemistry on polymorphism with a computational study, using the adsorption of tetracyanoquinodimethane (TCNQ) on Cu(001) and Ag(001) as an exemplary system.
The molecule TCNQ was chosen because it shows a strong covalent interaction with the Cu(001) substrate,\cite{tsengChargetransferinducedStructuralRearrangements2010} as well as significant charge transfer.\cite{torranceDifferenceMetallicInsulating1979}
Due to this strong interaction between adsorbate and substrate we expect the system to form commensurate adlayers, which simplifies the computational modeling.
Experimentally and computationally, TCNQ has previously been extensively studied on different noble metal substrates.\cite{tsengChargetransferinducedStructuralRearrangements2010,kamnaStrongElectronicPerturbation1998,sykesSubstrateMediatedInteractionsIntermolecular2003,
gonzalez-lakunzaFormationDispersiveHybrid2008,torrenteStructureElectronicConfiguration2008,yanAdsorptionTTFTCNQ2009,shiStructuralAnalysisElectronic2010,
martinezSimulatingOrganicmoleculeMetal2011,wackerlinAssembly2DIonic2011,barjaOrderedArraysMetal2013,fladischerHeliumAtomScattering2014,parkWeakCompetingInteractions2014,jeonThermodynamicControlTwoDimensional2016,
stradiUnderstandingSelfassemblyTCNQ2016, gerbertMolecularIonFormation2017,yamaneHighHoleMobilityMolecular2017,bloweyReevaluatingHowCharge2018,oteroComparativeComputationalStudy2019,
fuhr2DCuTCNQMetal2020,mousleyDirectExperimentalDetermination2023}
To isolate the effects of the lattice constant, we investigate how increasing the lattice constant of Cu by up to \qty{14.3}{\percent} (corresponding to the lattice constant of Ag) modifies the polymorphism of the adlayer.
These results are then compared to the polymorphism on Ag(001) to further distinguish between the effects that are caused by changes in lattice constant and surface chemistry.

\section{Methods}

\subsection{Computational Details}

All DFT calculations presented in this work were performed with the FHI-aims package,\cite{blumInitioMolecularSimulations2009}
employing the Perdew-Burke-Ernzerhof (PBE)\cite{perdewGeneralizedGradientApproximation1996} exchange-correlation functional
in conjunction with the non-local many-body dispersion correction method (MBD-NL).\cite{hermannDensityFunctionalModel2020}
"Tight" basis set defaults as provided by FHIaims (post 2020) were used for the C, N, and H species, while the "intermediate"
basis set defaults were used for the metal species Cu, Ag and Au.
To be consistent with our choice of functional, we determined the lattice constants of the metals computationally.
This yielded a lattice constant of \qty{3.598}{\angstrom} for Cu, \qty{4.114}{\angstrom} for Ag and
\qty{4.138}{\angstrom} for Au.
The surface slab was modeled with six layers for all metal substrates, with a vacuum region of
at least \qty{35}{\angstrom} in the $z$-direction. During geometry optimizations all metal atoms below the second
layer of the slab were kept fixed at their bulk position, as commonly done in literature.\cite{hofmannFirstprinciplesCalculationsHybrid2021}

For the SCF convergence we employed settings that required the total energy and the charge density
to change by less than \qty{e-6}{\electronvolt} and \qty{e-5}{} electrons between iterations, respectively.
For geometry relaxations we additionally required the forces to differ by less than \qty{e-4}{\electronvolt\per\angstrom} between iterations.
Geometry relaxations themselves were deemed converged once the total force on all atoms fell below \qty{0.01}{\electronvolt\per\angstrom}.
To occupy the Kohn-Sham eigenstates a Gaussian broadening scheme with a broadening width of \qty{0.01}{\electronvolt} was used.
The reciprocal space was sampled using a Generalized Monkhorst-Pack grid\cite{monkhorstSpecialPointsBrillouinzone1976,wisesaEfficientGenerationGeneralized2016}
with a $k$-point density equivalent to $36/b$ \unit{\angstrom^{-1}} in $x$ and $y$ direction and on $k$-point
in $z$ direction. Here $b$ is the primitive reciprocal lattice vector of the substrate in $x$/$y$ direction.
For geometry optimizations the $k$-point density was halved to keep the calculations tractable.

\section{Results and Discussion}

The formation of an organic adlayer on a metal substrate is governed by the interplay between adsorbate-substrate interaction and adsorbate-adsorbate interaction.
Since we want to properly distinguish between these two types of interactions, we make the following assumption:
Each polymorph can be constructed by taking the adsorption geometries of isolated adsorbates and arranging them into a close-packed layer.\cite{hormannSAMPLESurfaceStructure2019}
Put another way, the adsorption geometries of isolated adsorbates serve as the building blocks of all polymorphs.
It is important to note, that this assumption only holds if the adsorbate-adsorbate interaction is small enough that the isolated adsorption geometries are not significantly altered when put into a monolayer.
In particular, the adsorption geometries are required to be \enquote{locked} to their adsorption site and cannot be dislodged by the presence of other adsorbates.
This is generally the case when the adsorbate-substrate interaction for each adsorbate is much larger that the intermolecular interaction.\cite{braunNonlinearDynamicsFrenkel1998}

The adsorption energy, \(E_{\text{ads}}\), is defined as follows:
\begin{equation}\label{eqn:e_ads}
  E_{\text{ads}} = E_{\text{int}} - E_{\text{sub}} - N E_{\text{mol}},
\end{equation}
where \(E_{\text{int}}\) is the energy of the organic/metal interface, \(E_{\text{sub}}\) the energy of the clean substrate, \(N\) the number of molecules present in the adlayer unit cell and \(E_{\text{mol}}\) the energy of the molecule in gas phase.
The adsorbate-adsorbate interaction of a polymorph is then the difference between the adsorption energy of the polymorph and the adsorption energy of an isolated molecule.
With this split in mind, we first study the adsorbate-substrate interaction of a single TCNQ molecule under changing lattice constant and surface chemistry.
To this end, we explore the isolated adsorption geometries and their relative energies.
Subsequently, we construct representative polymorphs using these isolated adsorption geometries and investigate how the adsorbate-adsorbate interaction of these polymorphs is affected by lattice constant and surface chemistry.

\subsection{Adsorbate-Substrate Interaction}

To investigate how lattice constant and surface chemistry affect the adsorbate-substrate interaction, we perform a structure search for adsorption geometries of TCNQ on multiple substrates (for further details see the supporting information \Cref{sec:suppsearchdesc}).
We consider the Cu(001) substrate and the Ag(001) substrate to determine the influence of the surface chemistry, mainly the difference in reactivity between Cu and Ag.
In addition, we examine Cu(001) substrate with increased lattice constants.
We chose large changes of lattice constants, specifically, a 2\% and a 14.3\% increase, hereafter referred to as Cu-2\% and Cu-14\%, to clearly illustrate the effects at work.
We note that already the 2\% increase is inside the plastic deformation regime of Cu.\cite{rawatIntegratedExperimentalComputational2014}
The more extreme increase of 14.3\% was chosen to match the lattice constant of Ag and to properly distinguish between surface chemistry and lattice constant.
In addition, we expect that the massive increase in lattice constant will have a large impact on the adsorption geometries.

\begin{figure}[htb!]
  \centering
  \includegraphics[width=6in]{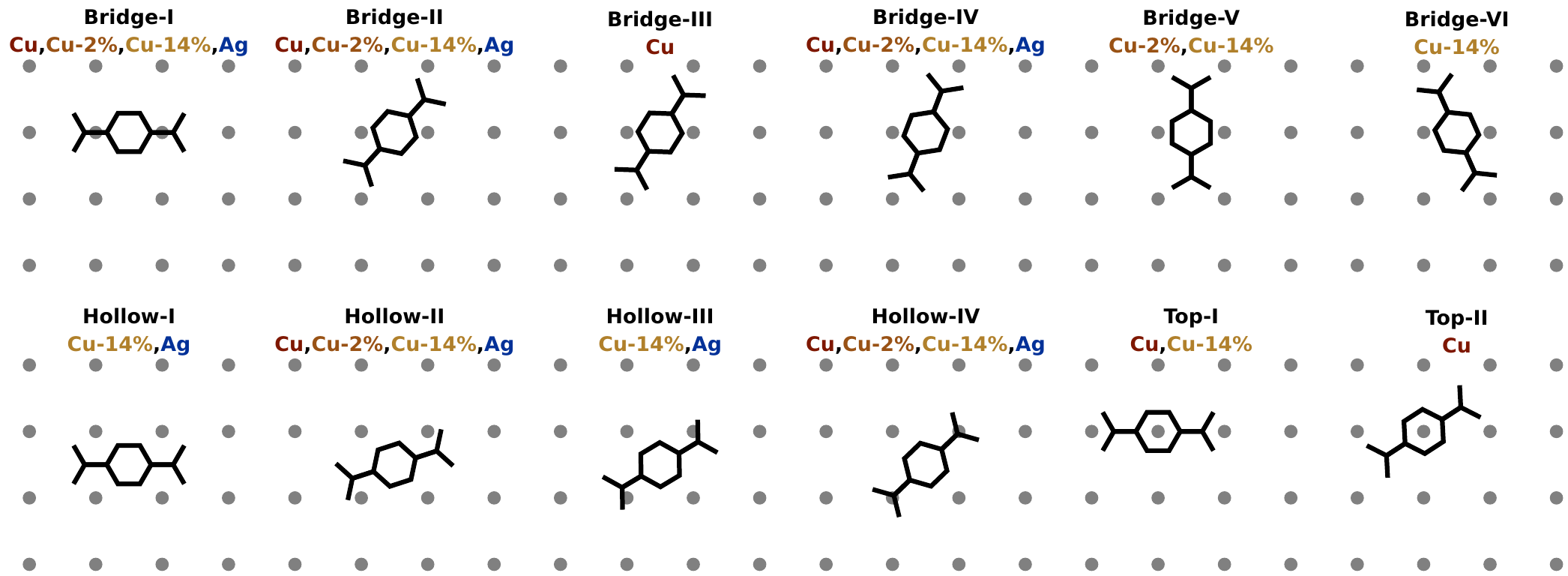}
  \caption{Schematic figure of the adsorption geometries of TCNQ to show all possible adsorption geometries.
  The gray points indicate a lattice with the lattice constant of Cu.
  More detailed depictions of the adsorption geometries can be found in the supporting information \Cref{sec:suppadsgeoms}.
  For the relative energies on the various substrates, see \Cref{fig:energy-comparison-2per} and \Cref{fig:energy-comparison-substrates}.}\label{fig:schematic-geometries}
\end{figure}

A schematic representation of all the (meta)-stable geometries can be found in \Cref{fig:schematic-geometries}, while more detailed visualizations of all calculated adsorption geometries can be found in the supporting information \Cref{sec:suppadsgeoms}.
We use the following nomenclature for the adsorption geometries: A geometry is named after the surface site on which the center of the phenyl ring of the TCNQ molecule is located, in this case either top, hollow or bridge.
For each site the geometries are enumerated with Roman numerals.
These numerals are ordered based on the angle between the long axis of the TCNQ molecule and the [100] direction of the substrate.
As \Cref{fig:schematic-geometries} shows, five adsorption geometries are consistently found on all substrates, that is, independent of the lattice constant or chemistry.
Two of these geometries, Bridge-I and Hollow-IV, adsorb in a symmetric fashion.
For instance, the hollow site shows a four-fold symmetry and rotating the adsorbate of the Hollow-IV geometry by 90\(^\circ\) results in the same adsorption geometry.
Since all substrates exhibit the same symmetries, it is not surprising that these symmetric geometries appear on all of them.
In contrast, the other geometries, Bridge-II, Bridge-IV and Hollow-II, do not satisfy any of the symmetry operations of the underlying substrate.

Next, we investigate how the adsorption geometries differ between the considered hypothetical substrates.
As one could expect, the Cu-2\% adsorption geometries include all the adsorption geometries from the standard Cu surface.
However, also a completely new adsorption geometry, Bridge-V, arises.
This shows that even comparatively small modifications of the substrate lattice constant can impact the single molecule adsorption.
The more extreme substrate modification Cu-14\% has less geometries in common with the standard Cu surface.
The Bridge-III and Top-II geometries are no longer stable after the significant increase in lattice constant.
At the same time, several new geometries appear.
These new geometries are Bridge-V, which already appears for the 2\% increase, Bridge-IV, Hollow-I and Hollow-III.

When comparing the Cu surface and Ag surface, we also find several differences in the adsorption geometries.
The Ag surface has more stable geometries on the hollow site, while adsorption on the top site becomes completely unfavorable.
In contrast, \emph{all} adsorption geometries found on Ag are also included in the set of adsorption geometries found on Cu-14\%.
In other words, the matching of the lattice constant leads to much more similar adsorption geometries.
Together, this paints the picture of the lattice constant of the substrate being more relevant for the adsorption geometry than the details of the chemistry.

Based on our building block approach, the presence or absence of adsorption geometries impacts the polymorphism, as they determine possible packing motifs.
In other words, the disappearance of adsorption geometries can exclude a whole class of polymorphs, while the emergence of new adsorption geometries can give rise to new polymorphs.
At the same time, the relative energetic ordering of the adsorption geometries is crucial for the polymorphism, as the energetically most favorable geometries are likely to occur in stable polymorphs.
For that reason, a comparison of the relative energies between substrates is necessary if we want to make more concrete statements about the polymorphism.

\begin{figure}[htb!]
  \centering
  \includegraphics{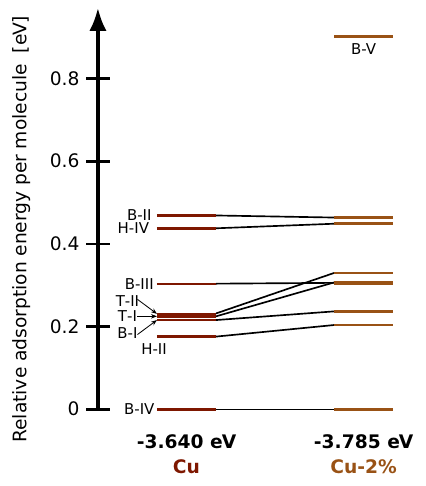}
  \caption{Relative energies of the adsorption geometries on the Cu surface and the Cu-2\% surface.
  The geometry names have been abbreviated; T for Top, B for Bridge and H for Hollow.}\label{fig:energy-comparison-2per}
\end{figure}

As a first step, we analyze the standard Cu surface and compare its relative adsorption energies to the Cu-2\% surface.
\Cref{fig:energy-comparison-2per} shows the relative energies of the adsorption geometries with respect to the energetically most favorable one for both Cu and Cu-2\%.
For Cu the best the best adsorption geometry is Bridge-IV, which is present on all surfaces.
There is a gap of nearly \qty{0.2}{\electronvolt} between the Bridge-IV geometry and the next best geometry, Hollow-II.
In addition, the Bridge-IV geometry also matches the lowest energy adsorption geometry found by Tseng et al.\cite{tsengChargetransferinducedStructuralRearrangements2010} on Cu(001) barring a small shift off of the bridge site.
The Cu-2\% surface shows similar relative energetics and has that same best adsorption geometry.
When comparing the surfaces, we observe that the change in relative energy for most geometries is rather small, not exceeding \qty{0.03}{\electronvolt}.
Only the top geometries, Top-I and Top-II, undergo a more significant shift in energy, suggesting that, for TCNQ, the top site is more dependent on the lattice constant than the other adsorption sites.
These two geometries are also the only ones where the energetic order of the adsorption geometries changes.
Moreover, the Bridge-V geometry that appears on Cu-2\% is energetically far above the other geometries.
Thus, despite being an additional \enquote{building block} for Cu-2\%, the Bridge-V geometry is unlikely to contribute to any stable polymorph on that substrate.

\begin{figure}[htb!]
  \centering
  \includegraphics{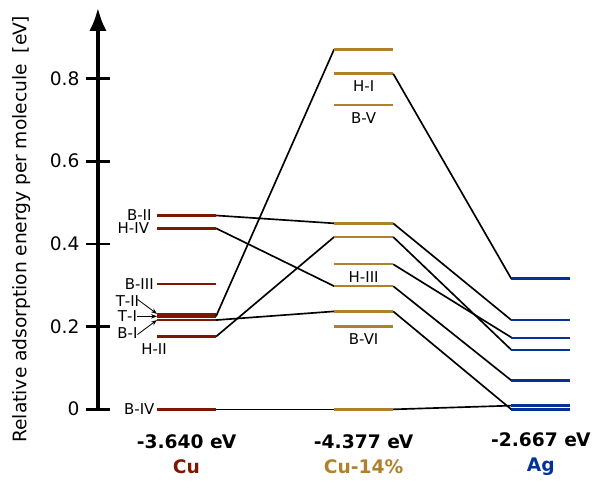}
  \caption{Relative energies of the adsorption geometries on the Cu surface and the Cu-2\% surface.
  The geometry names have been abbreviated; T for Top, B for Bridge and H for Hollow.}\label{fig:energy-comparison-substrates}
\end{figure}

Next, we compare Cu to Cu-14\% and Ag. The corresponding relative adsorption energies are visualized in \Cref{fig:energy-comparison-substrates}.
On the Cu-14\% surface, the sizable energetic gap between the Bridge-IV geometry and the other geometries persists.
Furthermore, we observe that the extreme increase in lattice constant can result in drastic change of the relative energy with respect to the Cu surface, as well as pronounced changes in the energetic ordering.
In particular, the Top-I geometry changes its relative energy by more than \qty{0.6}{\electronvolt}.
On the other hand, Bridge-II and Bridge-I hardly change their relative energy even with the large increase in lattice constant.
On the Ag surface the best geometry is the symmetric Bridge-I geometry.
However, it is only \qty{0.01}{\electronvolt} more favorable than Bridge-IV.
In comparison to the Cu surfaces, on Ag the adsorption geometries get closer to each other energetically.
We attribute this to the change in surface chemistry, in particular the stronger covalent bonds between the N atoms and metal atoms on the Cu surface.

To conclude this section, we discuss some implications for the polymorphism of these systems, based on our analysis of the adsorbate-substrate interaction.
Although different adsorption geometries appear, changing the lattice constant seems not to impact the polymorphism considerably, mainly due to the large energetic gap between the best adsorption geometry and the rest.
Still, the large changes in the energetic order for some of the adsorption geometries does make it plausible that, for another organic/metal interface, changing the lattice constant could change the best adsorption geometry or make other adsorption geometries more relevant energetically.
Changing the surface chemistry from Cu to Ag yields a different best adsorption geometry and brings the all geometries closer together in energy.
For that reason, we expect that adsorbate-adsorbate interaction will play a more important role for Ag, as it can more easily overcome the smaller gap in adsorbate-substrate interaction present on the Ag surface.

\subsection{Adsorbate-Adsorbate Interaction}

\begin{figure}[htb!]
  \centering
  \includegraphics{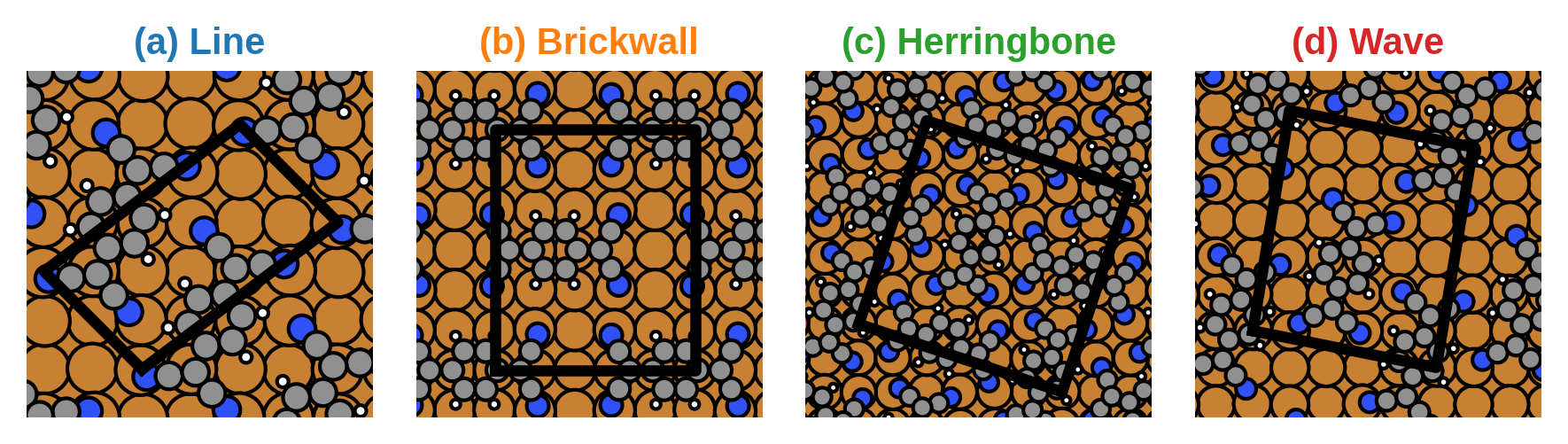}
  \caption{Top view of the polymorphs under consideration on Cu(001) before optimisation.
  (a) Line arrangement consists of the Bridge-IV geometry. (b) Brickwall polymorph build from Bridge-I geometries. (c) Herringbone packing consisting of the Hollow-II geometry. (d) Wave arrangement formed from the Hollow-II geometry.}\label{fig:polymorph-structures}
\end{figure}

In order to investigate the adsorbate-adsorbate interaction, we need to transition from single molecule adsorption to full molecular adlayers, which are constructed by arranging the adsorption geometries from the last section in specific packing motifs.
Since we assume that adsorbates are \enquote{locked} to the adsorption sites we found in the previous section, the intermolecular distances depend on the present single molecule adsorption geometries and the lattice constant of the underlying substrate.
In other words, the lattice constant directly affects the adsorbate-adsorbate interaction by fixing the distances between adsorbates and in turn the polymorphism.
In this section we want to directly observe this influence by considering four distinct polymorphs of TCNQ.
These polymorphs are depicted on the standard Cu surface in \Cref{fig:polymorph-structures}.
The first polymorph (\enquote{Line}) consists of diagonal lines formed by the Bridge-IV geometry.
It corresponds to the polymorph found by Tseng et al.\cite{tsengChargetransferinducedStructuralRearrangements2010}
The second polymorph (\enquote{Brickwall}) consists of Bridge-I geometries forming a brickwall pattern.
The third polymorph (\enquote{Herringbone}) also contains the Hollow-II geometry, but in a herringbone pattern.
The fourth polymorph (\enquote{Wave}) is formed from a wavelike arrangement of Hollow-II geometries.

Before we discuss the polymorphs on the substrate, we consider the free-standing monolayers of the polymorphs.
While free-standing monolayers do not include all the effects of the substrate, such as charge transfer into the adsorbates,they are cheap to calculate and allow us to explore the full range between the Cu lattice constant and the Ag lattice constant in more detail.
We construct the free-standing monolayers by taking the adlayers from the Cu substrate and scaling up the cell size by up to 14\%.
We ensured that for the adsorbates inside the cell, only the intermolecular distances are affected by the scaling process.

\begin{figure}[htb!]
  \centering
  \includegraphics{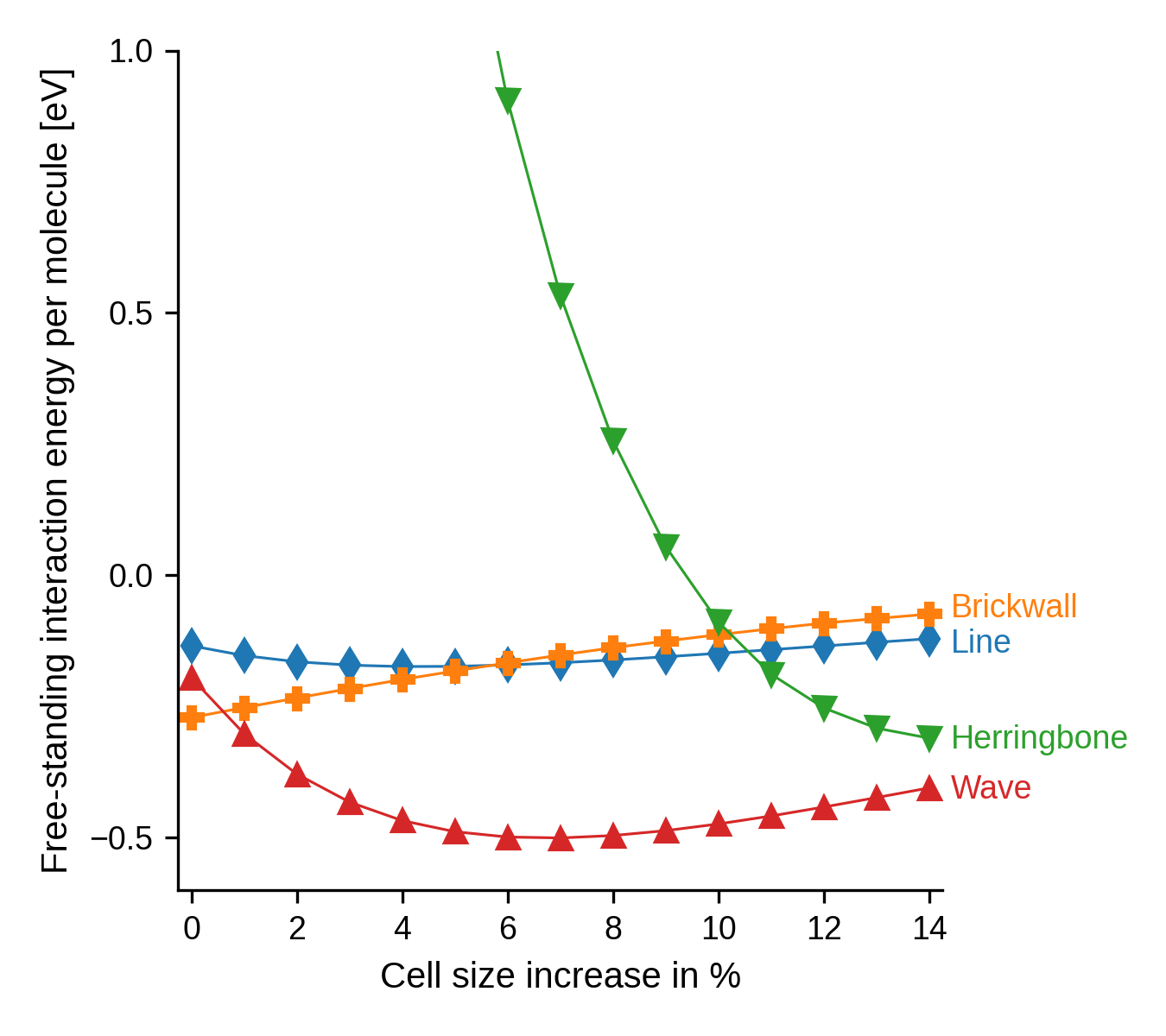}
  \caption{Change in gas-phase interaction energy per molecule for the polymorphs under consideration.
  The blue markers represent the Line polymorph, the orange markers the Brickwall polymorph, the green markers the Herringbone polymorph and the red markers the Wave polymorph.}\label{fig:polymorph-interaction-energy-gas-phase}
\end{figure}

The free-standing interaction energy for the scaled monolayers is displayed in \Cref{fig:polymorph-interaction-energy-gas-phase}.
The different interaction energy curves demonstrate that the intermolecular distance can have an impact on the interaction energy that varies from polymorph to polymorph.
The Line polymorph (shown in blue) shows attractive interaction energy over the complete considered range.
However, the interaction energy is barely affected when changing the lattice constant, that is, the intermolecular distance.
The Wave polymorph and the Brickwall polymorph also have purely attractive interaction.
In contrast, the Herringbone pattern shows strong repulsive interaction for the majority of the investigated region, only becoming attractive around a lattice constant increase of 10\%.
For that reason, we expect that in particular polymorphs with repulsive interactions are favored upon increasing the lattice constant.

Next we investigate the four polymorphs on the on the Cu, Cu-14\% and Ag substrates.
To obtain accurate energies, we performed geometry relaxations for all twelve systems.
As expected, for most polymorphs the optimization causes no significant changes in the geometry.
Only the herringbone polymorphs on Cu undergoes significant change.
While the adsorbates do not move from their adsorption sites, they buckle significantly.
This is a direct consequence of the repulsive interactions observed in gas phase.
Since the adsorbate are to close together on the Cu substrate and they cannot move from their adsorption site, they buckle in order to minimize the repulsive interaction between each other.

Calculating the adsorbate-adsorbate interaction on the substrate is not as straight-forward as in gas phase, particularly since we need to exclude the energetic contribution from the deformation of the adsorbate and substrate.
To this end we introduce the \enquote{modified adsorption energy}, \(\bar{E}_{\text{ads}}\), defined as follows
\begin{equation}\label{eqn:e_ads_mod}
  \bar{E}_{\text{ads}} = E_{\text{int}} -  E_{\text{sub,def}} - \sum_{i=1}^{N} E^{(i)}_{\text{mol,def}}.
\end{equation}
Here, \(E_{\text{sub,def}} \) is the energy of the deformed substrate from the interface without the adsorbates and \(E^{(i)}_{\text{mol,def}}\) is the energy of the deformed adsorbates in gas phase.
The modified adsorption energy is equivalent to taking the adsorption energy \(E_{\text{ads}}\) and subtracting the deformation energy of the substrate and the adsorbate.

To determine the adsorbate-adsorbate interaction, we first calculate the modified adsorption energy of all the adsorption geometries present in the given polymorph. We refer to these energies as \(\bar{E}^{(i)}_{\text{ads,geom}}\), where \(i\) enumerates the adsorption geometries.
To prevent interactions with periodic replicas of the adsorption geometries, we perform these calculations in a \(2\times2\times1\) supercell of the polymorph unit cell. Then we calculate the modified adsorption energy of polymorph, \(\bar{E}_{\text{ads,poly}}\), and define the adsorbate-adsorbate interaction as the following difference
\begin{equation}\label{eqn:e_ads_ads}
  E_{\text{ads-ads}} = \bar{E}_{\text{ads,poly}} - \sum_{i=1}^{N} \bar{E}^{(i)}_{\text{ads,geom}}.
\end{equation}

\begin{figure}[htb!]
  \centering
  \includegraphics{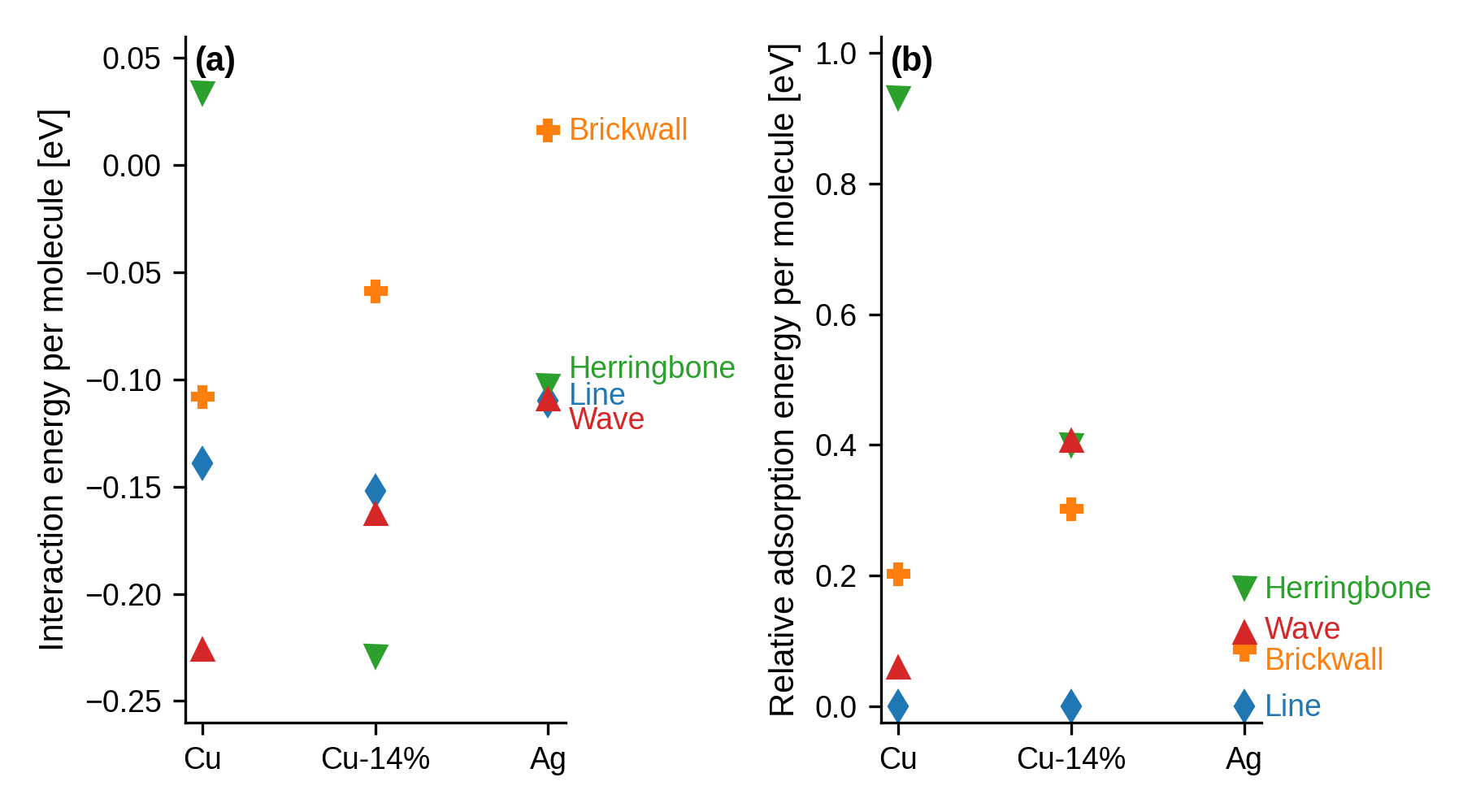}
  \caption{(a) Interaction energy of the four polymorphs on the different substrates. (b) Adsorption energies per molecule of the four polymorphs relative to the Line polymorph on the different substrates.
  The blue markers represent the Line polymorph, the orange markers the Brickwall polymorph, the green markers the Herringbone polymorph and the red markers the Wave polymorph.}\label{fig:polymorph-energies}
\end{figure}

The adsorption-adsorption energy, as defined in \Cref{eqn:e_ads_ads}, as well as the adsorption energies (\Cref{eqn:e_ads}) relative to the line polymorph are shown in \Cref{fig:polymorph-energies}, for the different substrates.
We observe some notable changes in the adsorbate-adsorbate interaction in \Cref{fig:polymorph-energies} (a).
Both the Brickwall and the Wave polymorph display interaction energies getting more repulsive as the substrate changes to Cu-14\% and Ag.
The Brickwall polymorph even has repulsive adsorbate-adsorbate interaction on Ag, despite the adsorbates being far apart from each other.
Furthermore, the adsorbate-adsorbate interaction of the Wave polymorph differs qualitatively from the free-standing interaction, as the interaction is more attractive for the standard Cu substrate, than for the Cu-14\% substrate.
The interaction of the Line polymorph undergoes less change than the other polymorph, staying between \qty{-0.10}{\electronvolt} and \qty{-0.15}{\electronvolt}.
The adsorbate-adsorbate interaction of the Herringbone polymorph changes from repulsive to attractive upon increasing the lattice constant by 14\%.
In addition, Herringbone shows the biggest change in interaction energy between Cu and Cu-14\%, amounting to nearly \qty{0.3}{\electronvolt}.
This large change is in qualitative agreement with the result of the free-standing monolayer, although less extreme, mainly because the repulsive interaction on Cu has been reduced by the strong buckling of the adsorbates.
Finally, we note that the adsorbate-adsorbate interaction becomes less attractive on Ag when compared to the Cu-14\% substrates, despite having the same intermolecular distances.
In other words, the underlying substrate and surface chemistry also affect the interaction between the adsorbates.

With the changes in adsorbate-adsorbate interaction in mind, we next discuss the relative energies of the polymorphs shown in \Cref{fig:polymorph-energies}.
The Line polymorph is energetically most favorable on all substrates.
On the Cu substrates, this can be attributed to the more favorable adsorbate-substrate interaction of the Bridge-IV geometry present in this polymorph.
The Wave polymorph has the more attractive adsorbate-adsorbate interaction on the standard Cu substrate, but only ranks second best.
On Ag the situation is different, as the adsorbate-substrate interaction is less important.
Here the Line polymorph is the best because it has more attractive adsorbate-adsorbate interaction than the Brickwall polymorph, as both polymorphs have similar adsorbate-substrate interaction.
Lastly, as expected from the adsorbate-adsorbate interaction and strong deformation, the Herringbone polymorph has the worst energy on Cu but significantly closes the gap on the surfaces with the larger lattice constant.

\section{Conclusion}

In this work, we examined how the geometry of the underlying substrate, in particular the lattice constant, can affect the polymorphism of organic/metal interfaces.
Using the adsorption of TCNQ on Cu(001) and Ag(001) as an example and also including two artificial Cu(001) substrate with increased lattice constants, we investigated how the lattice constant and the surface chemistry affect the adsorbate-substrate and the adsorbate-adsorbate interaction of organic/metal interfaces.

By performing structure search for the four different substrates, we discovered that these substrates have many adsorption geometries in common.
Nonetheless, the relative energetic order for some of the can quite substantially influenced if the lattice constant is increase sufficiently.
While the 2\% lattice constant increase does not strongly affects the adsorption energetics, it does cause a new adsorption geometry, albeit with unfavorable adsorption energy, to appear.
In addition, we observed a similarity between the adsorption geometries on Ag(001) and Cu(001) with Ag lattice constant.
This suggest that the lattice constant of the underlying substrate also impacts the structure of the adsorption geometries.

Afterwards, we investigated the adsorbate-adsorbate interaction by considering several polymorphs, including a polymorph that was found experimentally.
We observed that particularly close-packing polymorphs can massively lower their adsorbate-adsorbate interaction, when transitioning from the repulsive to the attractive regime.
However, even such close-packing polymorphs will not necessarily become stable, unless the adsorbate-substrate interaction changes favorably under the lattice constant increase.
Despite this complication, the potentially large gains in adsorbate-adsorbate interaction could still enable lattice-constant-induced phase transition for organic/metal interfaces, particularly those with adsorption geometries that are lie close to each other energetically.

\section{Acknowledgments}

Funding through the HI-TEq project of the Austrain Science Fund (FWF): Grant No. I5170-N (\url{https://doi.org/10.55776/I5170}) is gratefully acknowledged.
Computational results have been achieved using Austrian Scientific Computing (ASC).
\clearpage
\printbibliography

% ====== Supporting Information ======
\clearpage
%! TEX root = ./main.tex
\setcounter{page}{1}
\setcounter{secnumdepth}{1}
\renewcommand\thesection{S\arabic{section}}
\renewcommand{\thefigure}{S\arabic{figure}}
\renewcommand{\theequation}{S\arabic{equation}}
\setcounter{figure}{0}
\setcounter{equation}{0}

\begin{center}
  {\LARGE \bfseries {\Huge Supporting Information}\\ for\\ \enquote{Impact of the Lattice Constant on the Polymorphism of Organic/Inorganic Interfaces} \par}
  \vspace{1.2em}

  {\large
    Christoph Wachter\textsuperscript{1} and Oliver T. Hofmann\textsuperscript{1,*}
    \par
  }
  \vspace{0.6em}
  {\itshape
    \textsuperscript{1}Institute of Solid State Physics, NAWI Graz, Graz University of Technology, Petersgasse 16, 8010 Graz, Austria
    \par
  }
  \vspace{0.8em}
  {\small *Email: o.hofmann@tugraz.at \par}
\end{center}

\section{Structure Search Details}\label{sec:suppsearchdesc}

\begin{figure}[htb!]
  \centering
  \includegraphics[scale=0.5]{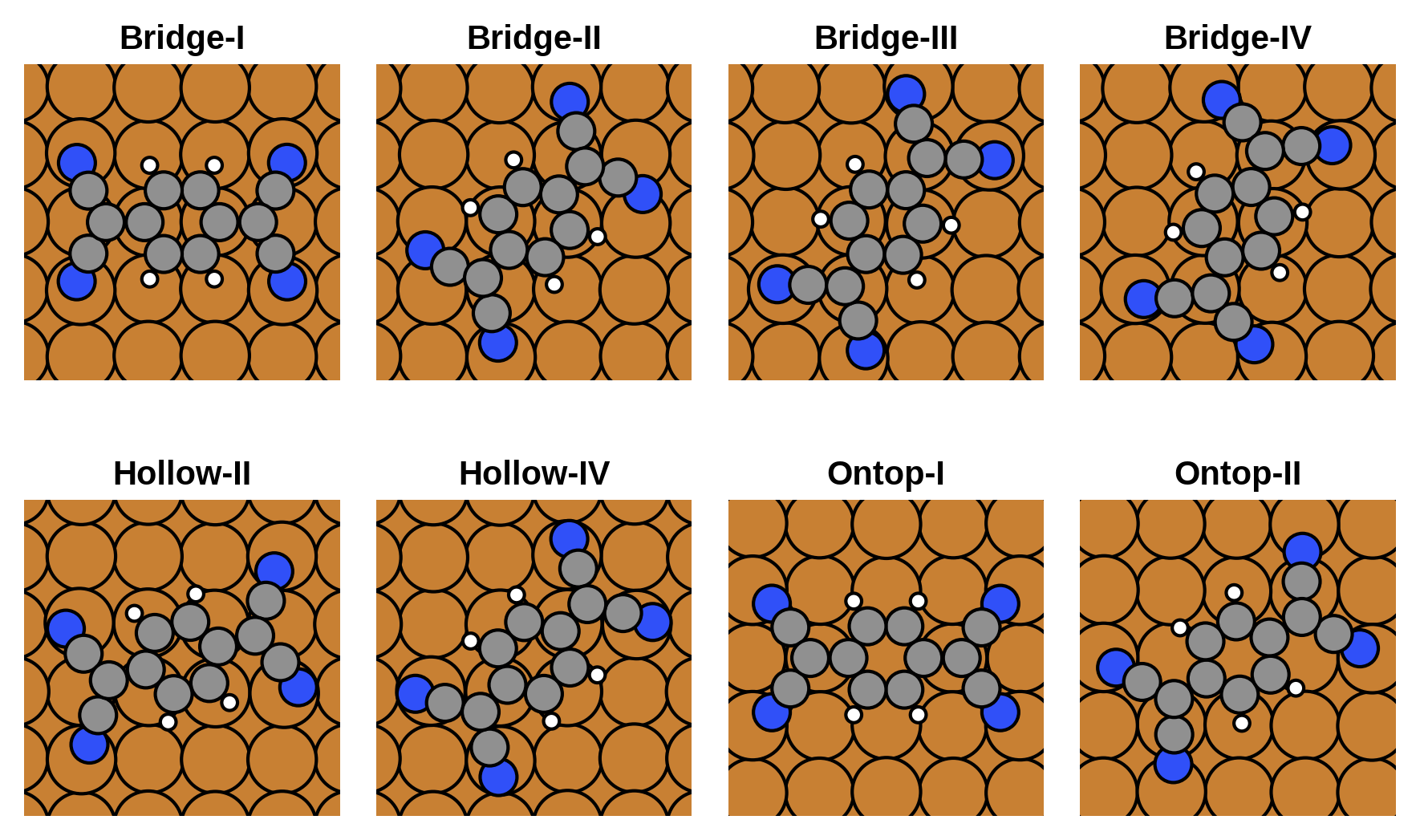}
  \caption{Top view of the adsorption geometries on the Cu(001) substrate.}\label{fig:cu-geoms}
\end{figure}

\begin{figure}[htb!]
  \centering
  \includegraphics[scale=0.5]{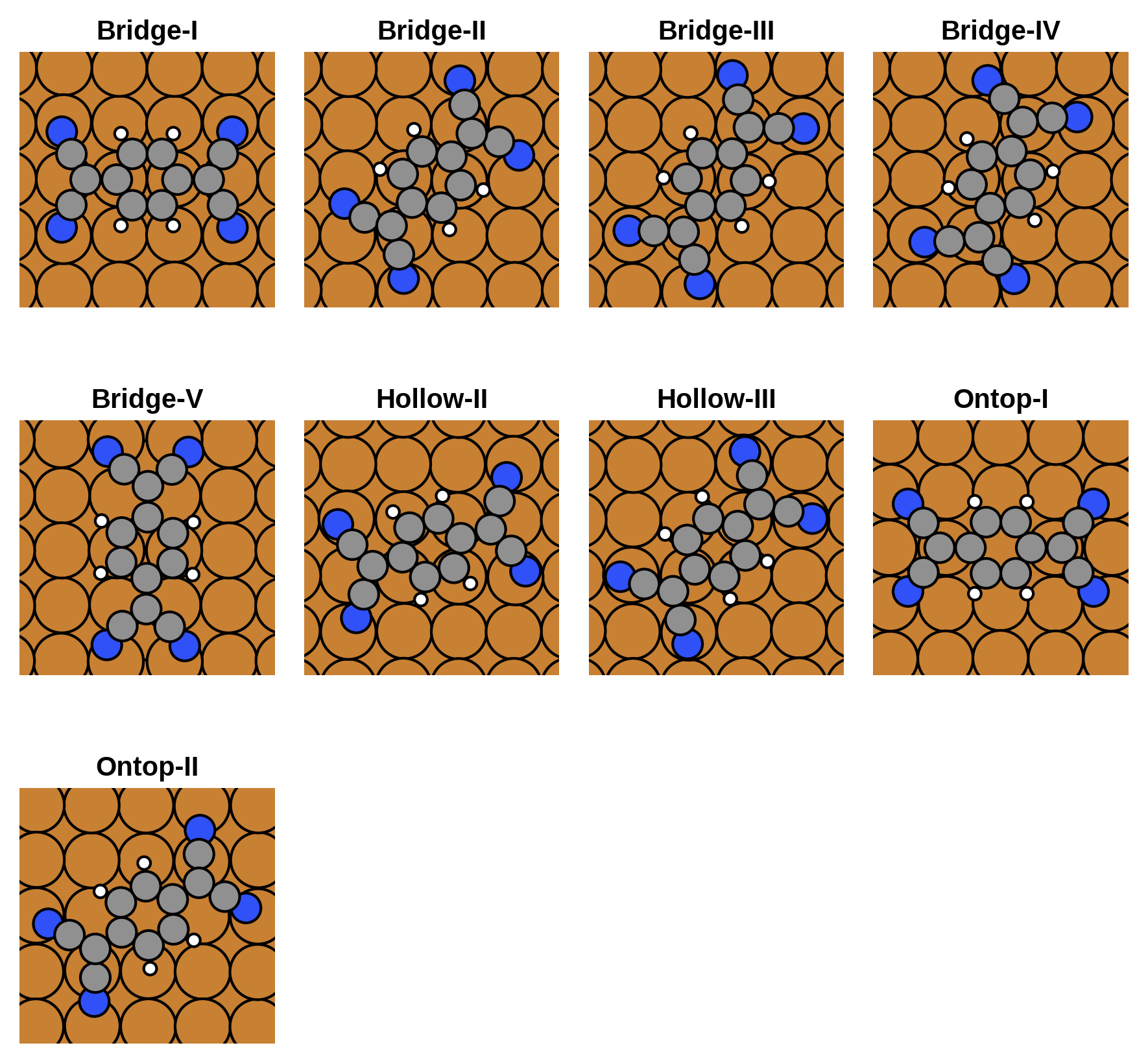}
  \caption{Top view of the adsorption geometries on the Cu(001) substrate with 2\% lattice constant increase.}\label{fig:cu-2-geoms}
\end{figure}

\begin{figure}[htb!]
  \centering
  \includegraphics[scale=0.5]{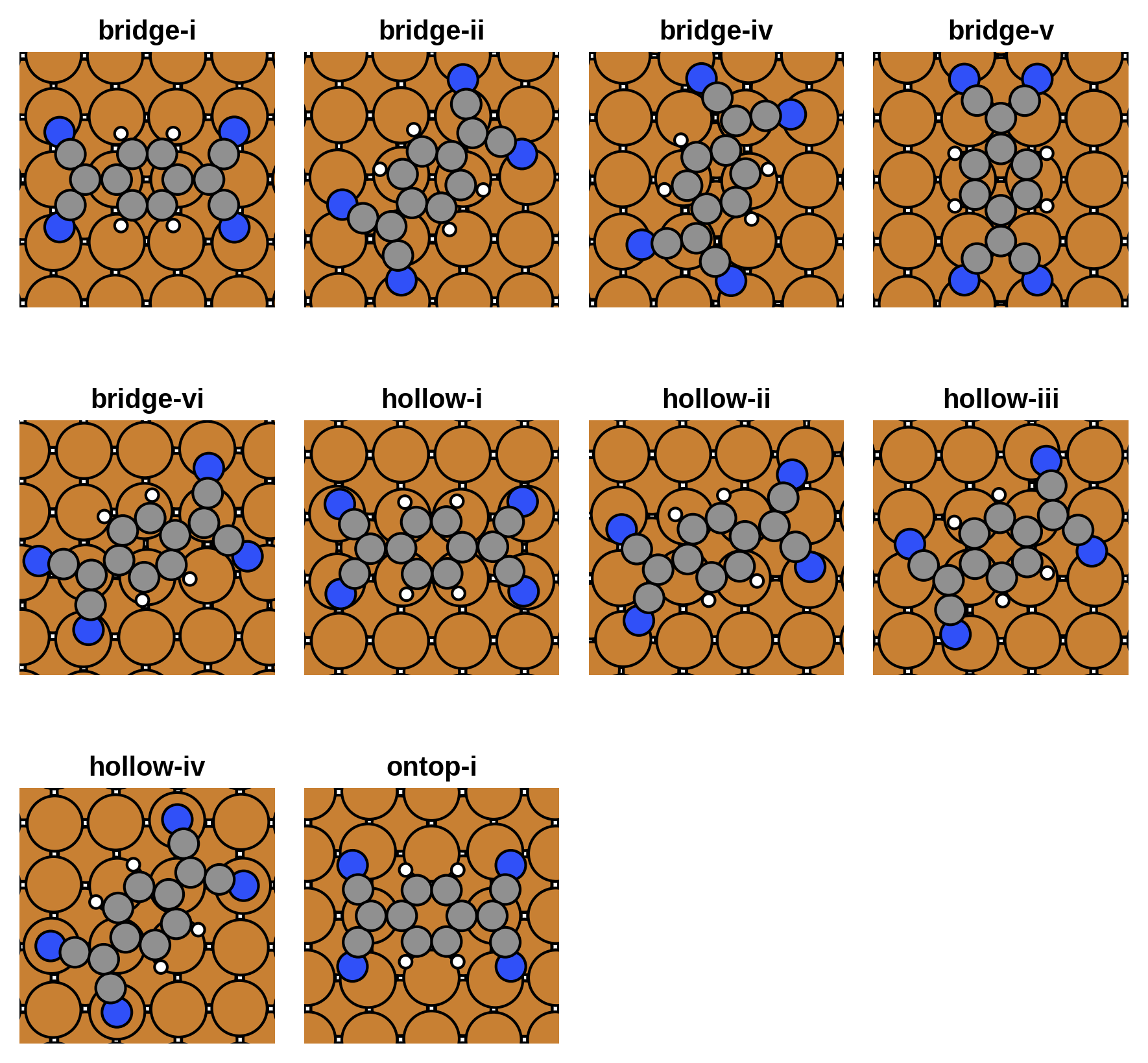}
  \caption{Top view of the adsorption geometries on the Cu(001) substrate with 14\% lattice constant increase.}\label{fig:cu-14-geoms}
\end{figure}

\begin{figure}[htb!]
  \centering
  \includegraphics[scale=0.5]{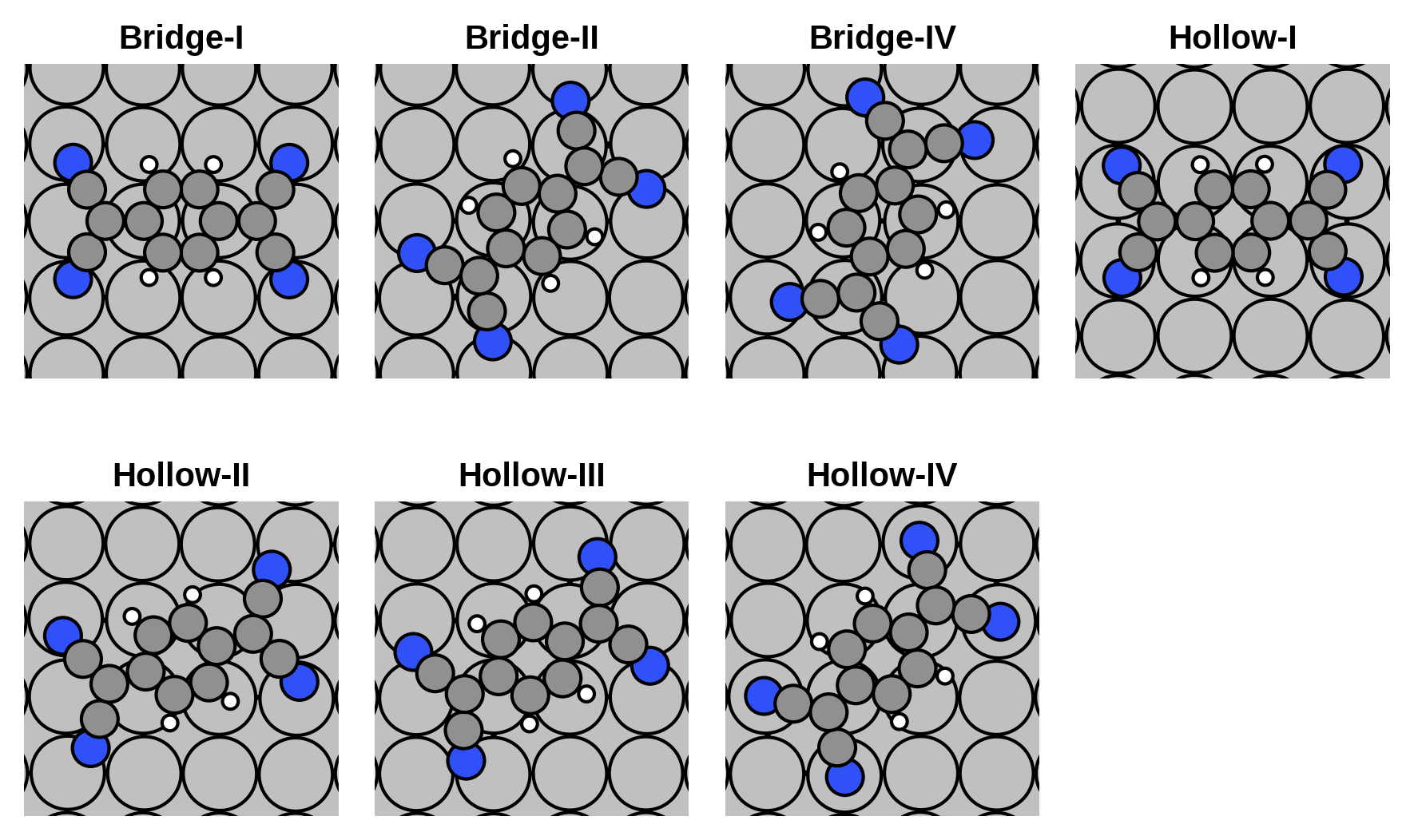}
  \caption{Top view of the adsorption geometries on the Ag(001) substrate.}\label{fig:ag-geoms}
\end{figure}

To determine the adsorption geometries we placed the adsorbate on different starting positions and performed geometry optimizations for all of them.
The starting positions were chosen as follows:
The adsorbate was placed on the top, hollow and bridge sites in various angles relative to the [100] direction.
For the top and hollow site we chose the angles of \qty{0}{\degree}, \qty{15}{\degree}, \qty{30}{\degree} and \qty{45}{\degree}.
For the bridge site we used the same angles as on the other two sites as well as \qty{60}{\degree}, \qty{75}{\degree} and \qty{90}{\degree}.
All calculations were performed in a \(6\times6\) supercell of the substrate to ensure that the adsorbates do not interact with their periodic replicas.
The initial adsorption height was the same for all starting points at \qty{2.1}{\angstrom}.
Once the optimizations were completed, symmetrically equivalent geometries were filtered, leaving those depicted in \Cref{fig:schematic-geometries} and \Cref{fig:cu-geoms,fig:cu-2-geoms,fig:cu-14-geoms,fig:ag-geoms}.

\section{Visualizations of all Adsorption Geometries}\label{sec:suppadsgeoms}

\end{document}